\documentstyle[aps,prl,multicol,epsf]{revtex}

\newcommand{\beq}{\begin{equation}}
\newcommand{\eeq}{\end{equation}}
\newcommand{\beqa}{\begin{eqnarray}}
\newcommand{\eeqa}{\end{eqnarray}}
\newcommand{\non}{\nonumber}

\begin{document}


\title{Coarsening Dynamics of a Nonconserved Field Advected by a Uniform 
Shear Flow}

\author{
Alan J. Bray  
and 
Andrea Cavagna 
}

\address{
Department of Physics and Astronomy, The University, Manchester, M13 9PL, 
UK}

\date{\today}

\maketitle

\begin{abstract}
We consider the ordering kinetics of a nonconserved scalar field advected by 
a uniform shear flow. Using the Ohta-Jasnow-Kawasaki approximation, 
modified to allow for shear-induced anisotropy, we calculate the asymptotic 
time dependence of the characteristic length scales, $L_\parallel$ and 
$L_\perp$, that describe the growth of order parallel and perpendicular to 
the mean domain orientation. In space dimension $d=3$ we find 
$L_\parallel \sim \gamma t^{3/2}$, $L_\perp \sim t^{1/2}$, where $\gamma$ 
is the shear rate, while for $d = 2$ we find 
$L_\parallel \sim \gamma^{1/2} t(\ln t)^{1/4}$, 
$L_\perp \sim \gamma^{-1/2}(\ln t)^{-1/4}$ . Our predictions for $d=2$ 
can be tested by experiments on twisted nematic liquid crystals.  
\end{abstract}


\begin{multicols}{2}

The coarsening dynamics of systems quenched from a disordered phase into 
a two-phase region are by now reasonably well understood 
\cite{Review}. Domains of the ordered phases form rapidly, and then coarsen 
with time $t$, i.e.\  there is a characteristic length scale (`domain size'), 
$L(t)$, which grows with time, typically as a power law.  Furthermore, 
there is good evidence for a form of dynamical scaling in which the 
domain morphology is statistically scale-invariant when all lengths are 
scaled by $L(t)$ \cite{Review}.

In recent years attention has been directed at systems subjected to external 
driving, e.g.\ an imposed shear \cite{OnukiRev,Beysens,Rothman}. 
The shear induces an anisotropy, leading to different 
coarsening rates in directions parallel and perpendicular to the flow.  
This has been observed in experiments \cite{Beysens}.

An important open question for such driven systems is whether the coarsening 
continues indefinitely (for an infinite system), as in the case of no 
driving, or whether the driving force arrests the coarsening, leading to a 
steady state. For the case of a sheared phase-separating binary fluid, it has
been argued, on the basis of the stability of a single drop of one fluid 
immersed in another, that the domain scale will eventually saturate at a 
maximum length scale, $L_{max}$, determined by the shear rate: 
$L_{max} \sim \sigma/\gamma\eta$, where $\sigma$ and $\eta$ are the surface 
tension and viscosity respectively \cite{OnukiRev}. In the multi-domain 
context, it has been argued \cite{Ohta} that a steady state would be 
achieved  through the shear-induced stretching and breaking of domains.  
However, the experimental evidence for a steady state is not completely 
clear-cut. In particular, as emphasized by Cates {\em et al.} \cite{Cates}, 
saturation of the domain length occurs naturally in a finite-size system 
when the domain length becomes of the same order as the system size.

A second important question concerns the nature of the growth laws, and the 
nature of scaling -- if it exists -- in this anisotropic system. Naively, 
one might expect two characteristic length scales, $L_\parallel$ and 
$L_\perp$, measuring correlations parallel and perpendicular to the flow. 
In dimension $d=2$, for example, it would be natural to conjecture that 
the coarsening follows conventional scaling when lengths along and 
perpendicular to the flow are scaled by these characteristic lengths. 
We will show, however, that the actual scaling is subtly different.

To address these issues it would be helpful to have an exactly soluble model. 
The large-$n$ limit of the $n$-vector model has been solved for a conserved 
order-parameter field, but without hydrodynamics (i.e.\ the order parameter 
is simply advected by the shear flow) \cite{Rapapa}. Length scales 
$L_\parallel$ and $L_\perp$, growing as $\gamma(t^5/\ln t)^{1/4}$ and 
$(t/\ln t)^{1/4}$ respectively, describe correlations along and perpendicular 
to the flow. Scaling is not strictly satisfied (there is instead 
a form of `multiscaling'), but this is presumably an artifact of the large-$n$ 
limit, as in the zero-shear case \cite{CZBH}. There is  {\em no} 
saturation at late times in this model.  However, for the large-$n$ 
model there are no domains, so the concept of stretching and breaking 
loses its meaning. What is needed is an analytically tractable model for 
a {\em scalar} order parameter. For a {\em nonconserved} field, the 
Ohta-Jasnow-Kawasaki (OJK) approximation \cite{OJK} fulfills this 
requirement: In the absence of shear, it captures the essential features 
of the coarsening process \cite{Review}.

In this Letter we present the results of applying the OJK approach to a 
nonconserved scalar field advected by a uniform shear flow. The equation 
of motion for the order parameter is 
\beq
\partial_t\phi + \nabla\cdot({\bf u}\phi) = \nabla^2 \phi - V'(\phi), 
\label{modelA}
\eeq
where $V(\phi)$ is a symmetric double well potential, and 
${\bf u} = \gamma y {\bf \hat{x}}$ is the velocity of the imposed shear flow, 
with the flow in the $x$-direction. Two aspects of real binary fluids are 
neglected in this model: the order parameter is not conserved, and it is 
simply advected by the shear rather than being coupled to the fluid velocity 
through the Navier-Stokes equation. This approach is, however, a very  
instructive first step which yields important insights concerning the 
main questions raised earlier, namely the nature the asymptotics and of 
dynamical scaling, in the physically correct context of a scalar field. 
Furthermore, in $d=2$ this model describes the coarsening dynamics of a 
twisted nematic liquid crystal, in which disclination lines, separating 
domains of opposite twist, relax viscously, driven by their line tension 
\cite{Orihara}. Under shear, this system will furnish an experimental test 
of our predictions. It should be noted that our analysis leads to very 
different behavior in $d=2$ and $d=3$: This is one of the main results of 
the present work.

The result of the OJK analysis is that, for space dimension $d = 3$, 
$L_\parallel \sim \gamma t^{3/2}$ and $L_\perp \sim t^{1/2}$, i.e.\ the 
coarsening rate parallel to the flow is enhanced by a factor $\gamma t$ 
relative to the unsheared system, while the growth of $L_\perp$ is unchanged  
(the same features describe the large-$n$ result \cite{Rapapa}). 
Hence there is no saturation of the coarsening. For $d=3$, furthermore, 
conventional scaling holds.  For $d=2$, however, very different results are 
obtained:  $L_\parallel \sim \gamma^{1/2} t(\ln t)^{1/4}$, and 
$L_\perp \sim \gamma^{-1/2}(\ln t)^{-1/4}$. These results 
imply $L_\parallel L_\perp \sim t$ for $d=2$, i.e.\ the product of the two 
length scales, or `scale area', is independent of the shear rate and has 
the same form as for the unsheared system, where 
$L_\parallel = L_\perp \sim t^{1/2}$. We will show that this result 
can be understood by a topological argument. An important subtlety in 
$d=2$ is that $L_\parallel$ and $L_\perp$ have to be defined as 
characteristic scales parallel and perpendicular to the mean domain 
orientation, instead of the flow direction. This distinction is not 
important for $d=3$, but crucial for $d=2$. Only with the new definition 
is dynamical scaling recovered.  For $d=2$, furthermore,  $L_\perp$ 
{\em decreases} with time asymptotically. Our approach breaks down, 
however, when $L_\perp$ becomes comparable with the width, $\xi$, 
of the interfaces between domains, which occurs after a time of order 
$\exp({\rm const}/\gamma^2\xi^4)$, when we expect the domains to break, 
as observed in simulations with conserved dynamics in $d=2$ \cite{Ohta}.

The OJK approach starts from the Allen-Cahn equation \cite{AC} relating 
the normal component of the interface velocity, $v_n$, to the local 
curvature of the interface, $K=\nabla\cdot{\bf n}$, where ${\bf n}$ is 
the normal to the interface,
\beq
v_n = -\nabla\cdot{\bf n} + {\bf u}\cdot{\bf n},
\label{AC}
\eeq
where the final term is the drift due to the shear. The derivation of this 
equation from (\ref{modelA}) follows the same route as the zero-shear case 
\cite{Review}. The next step is to introduce a smooth auxiliary field 
$m({\bf x},t)$ whose zeroes coincide with those of $\phi$. In a frame 
comoving with the interface one has $dm/dt = 0 = \partial_t m + 
v_n|\nabla m|$. Combining this with (\ref{AC}), and using 
\beq
{\bf n} = \nabla m/|\nabla m|,
\label{n}
\eeq 
yields the following equation for $m$,
\beq
\partial_t m + {\bf u}\cdot\nabla m = \nabla^2 m 
- \sum_{a,b=1}^d n_an_b\partial_a\partial_bm\ .
\label{OJK}
\eeq

So far this is exact. Equation (\ref{OJK}) is highly non-linear, however, 
due to the implicit dependence of ${\bf n}$ on $m$ through (\ref{n}). 
The OJK approximation involves linearizing the $m$ equation by replacing 
the product $n_an_b$ by its spatial average,
\beq
D_{ab}(t) = \langle n_a({\bf x},t)\,n_b({\bf x},t)\rangle\ ,
\label{D}
\eeq 
leading to the following equations for $m$ and $D$:
\beqa
\label{mOJK}
\partial_t m + \gamma y\,\partial_xm & = & \nabla^2 m 
                   - D_{ab}(t) \partial_a \partial_b m  \\
D_{ab} & = & \left \langle \frac{\partial_a m\,\partial_b m}
{(\nabla m)^2} \right\rangle\ .
\label{DOJK}
\eeqa

In the absence of shear the coarsening is isotropic, and the matrix $D$ 
has the simple form $D_{ab}=\delta_{ab}/d$, independent of $t$. The resulting 
diffusion equation for $m$ is readily solved, leading to a $\sqrt{t}$ 
coarsening.  For $\gamma \ne 0$, the coarsening is anisotropic: $D_{ab}$ 
is both  non-diagonal and time-dependent. From the definition (\ref{D}) 
of $D_{ab}$, though, the sum-rule, ${\rm Tr}\,D = 1$, is trivially valid. 
From the symmetry of (\ref{mOJK}) under the combined 
transformations $x \to -x$, $y \to -y$ at fixed $z$, and under the 
separate transformation $z \to -z$ at fixed $x,y$, we see that $D_{ab}$ 
has a block diagonal form, with $D_{xy} = D_{yx}$ the only non-zero 
off-diagonal elements.

In Fourier space, (\ref{mOJK}) reads
\beqa
\label{Fourier}
\frac{\partial m({\bf k},t)}{\partial  t}-
\gamma k_x\frac{\partial m({\bf k},t)}{\partial k_y} & = & 
-\sum_{ab}\Omega_{ab}(t)\,k_a k_b\,m({\bf k},t)\ , \\
\Omega_{ab}(t) & = & \delta_{ab} - D_{ab}(t)\ .
\label{Omega}
\eeqa
This is readily solved by the change of variables 
${\bf q}= A{\bf k}$, $\tau = t$, and $\mu({\bf q},\tau) = m({\bf k},t)$, 
where $A$ has elements 
\beq
A_{ab} = \delta_{ab} + \gamma t\,\delta_{a2}\delta_{b1}.
\label{A}
\eeq 
In the new variables, the left-hand side of (\ref{Fourier}) becomes 
$\partial_\tau \mu({\bf q},\tau)$. Integrating the equation, and 
transforming back to the original variables, gives
\beq
m({\bf k},t) = m(\tilde{\bf k}(t),0)
\exp\left(-\frac{1}{4}\sum_{ab} k_a M_{ab}(t)\,k_b \right),
\label{m}
\eeq
where $\tilde{\bf k}(t) = (k_x,k_y+\gamma k_x t,k_z)$ and the matrix 
$M$ is given by
\beqa
\label{M}
M(t) & = & A^T(t)\,R(t)\,A(t) \\
R(t) & = & 4\,\int_0^t dt'\,[A^T(t')]^{-1}\Omega(t')[A(t')]^{-1}. 
\label{R}
\eeqa

Equations (\ref{m}--\ref{R}) determine the function $m({\bf k},t)$ 
completely if the matrix $D_{ab}(t)$ is known. However, $D$ is itself 
determined from the distribution for $m$, via (\ref{DOJK}), so we have 
to solve these equations self-consistently. We take the initial condition, 
$m({\bf k},0)$ to be a gaussian random variable with correlator   
$\langle m({\bf k},0)m({\bf k}',0) \rangle = 
\Delta\,\delta({\bf k}+{\bf k}')$. Then the real-space correlation function 
of $m$, $G({\bf r},t) = \langle m({\bf x}+{\bf r},t)m({\bf x},t)\rangle$
is obtained from (\ref{m}) as 
\beq
G({\bf r},t) = G(0,t)\,
\exp\left(-\frac{1}{2}\sum_{a,b}r_a (M^{-1})_{ab}\,r_b\right)\ ,
\label{G}
\eeq
where the precise expression for $G(0,t)$ is not relevant in what follows. 
All the information concerning domain growth in this system is contained 
in the matrix $M_{ab}(t)$.

To obtain a closed set of equations we have to express the elements of the 
matrix $D$ in terms of the elements of $M$. To to this we first rewrite 
(\ref{DOJK}) using an integral representation for the denominator: 
\beq
D_{ab} = \frac{1}{2}\int_0^\infty du\,\left\langle\partial_am\,\partial_bm\,
\exp\left(-\frac{u}{2}(\nabla m)^2\right)\right\rangle\ .
\label{Dint}
\eeq
Since $m$ is a Gaussian field, the required average can be computed using
the probability distribution $P({\bf v})$ of the vector ${\bf v} = \nabla m$. 
This distribution is determined by the correlator
$\langle v_a\,v_b\rangle = -\partial_a\partial_b\, 
\langle m({\bf r},t)\,m({\bf 0},t) \rangle|_{{\bf r} = {\bf 0}}
 = \langle m^2 \rangle\, (M^{-1})_{ab}$, 
from which one infers that
\beq
P({\bf v}) \propto \exp\left(-\frac{1}{2\langle m^2 \rangle}\,
\sum_{a,b} v_a\,M_{ab}\,v_b\right)\ .
\eeq
Carrying out the average in (\ref{Dint}) gives
\beqa
\label{Dint1}
D_{ab} & = & 
\frac{1}{2}\int_0^\infty du\,\left(\frac{\det M}
{\det N}\right)^{1/2}\,
(N^{-1})_{ab} \\
N_{ab} & = & M_{ab} + u\,\delta_{ab}\ .
\label{Dint2}
\eeqa

We now proceed to outline the solution of this closed set of equations. 
Full details will be given elsewhere \cite{CB}. We are interested in the 
large-$t$ asymptotics. This limit simplifies the analysis which is still, 
however, quite subtle. The results are very different in three and two 
dimensions, so we discuss these cases separately.

We begin with some general remarks on the expected form of $D_{ab}$. The 
effect of the shear is to produce elongated domain structures aligned, at 
late times, at a small angle ($\sim 1/\gamma t$ - see below) to the flow 
direction. As a result, the component $n_1$ of the normal to the interface 
is very small almost everywhere at late times, implying $D_{11} \to 0$ for 
$t \to \infty$. The sum rule, ${\rm Tr}\,D=1$, is therefore exhausted by 
the remaining diagonal components of $D$ for large $t$. In particular, 
for $d=2$ we have $D_{22} \to 1$ for $t \to \infty$, and 
$\Omega_{11} \to 0$ in (\ref{Fourier}). This case requires special care. 
For $d=3$, on the other hand, $D_{22} + D_{33} \to 1$, and 
it turns out that both $D_{22}$ and $D_{33}$ approach non-zero limits. 
This case is, therefore, simpler to analyse.

{\bf The case} ${\bf d=3}$.  With the assumption that $D_{11} \to 0$ 
and $D_{12} \sim 1/t$, while $D_{22}$ and $D_{33}$ remain non-zero for 
$t \to \infty$, the asymptotics of the matrix elements $M_{ab}$ are 
readily obtained from (\ref{Omega}), (\ref{A}), (\ref{M}) and (\ref{R}):
\beqa
M_{11} & = & \frac{4}{3}\gamma^2t^3\,(1-D_{22}^\infty) \non \\
M_{12} & = & 2\gamma t^2\,(1-D_{22}^\infty) \non \\
M_{22} & = & 4t\,(1-D_{22}^\infty) \non \\
M_{33} & = & 4t\,D_{22}^\infty\ ,
\label{MK}
\eeqa
where $D_{22}^\infty$ is the large-$t$ limit of $D_{22}$, 
while $M_{13}=M_{23}=0$ by symmetry.

Using these limiting forms, the 
integrals (\ref{Dint1}) can be evaluated asymptotically. After some 
algebra one finds
\beq
D_{22}^\infty = \left(1 + \frac{1}{2}\,
\left(\frac{1-D_{22}^\infty}{D_{22}^\infty}\right)^{1/2}\right)^{-1} 
\label{D22}
\eeq
while $D_{33}^\infty = 1 - D_{22}^\infty$ and $D_{13}=D_{23}=0$. 
Equation (\ref{D22}) has the non-trivial solution $D_{22}^\infty = 4/5$, 
implying $D_{33}^\infty=1/5$ and demonstrating the self-consistency of 
the initial ansatz. 
Finally we find $D_{11} \to 3\ln(\gamma t)/(\gamma t)^2$ and 
$D_{12} \to -6/(5\gamma t)$, consistent with our initial assumption.

The characteristic length scales in the system are given, from (\ref{G}), 
by the square roots of the eigenvalues of the matrix $M$. Using (\ref{MK}), 
with $D_{22}^\infty = 4/5$, we find
\beq
L_\parallel = \frac{2}{\sqrt{15}}\,\gamma t^{3/2},\ \ 
L_\perp = \frac {1}{\sqrt{5}}\,t^{1/2},\ \ 
L_3 = \frac{4}{\sqrt{5}}\,t^{1/2},
\eeq
for $t \to \infty$. The corresponding eigenvectors are 
\beq
{\bf e}_\parallel = 
\left(\begin{array}{c} 1 \\ \frac{3}{2\gamma t} \\ 0 \\ \end{array}\right),
\ \ {\bf e}_\perp = 
\left(\begin{array}{c} -\frac{3}{2\gamma t} \\ 1 \\ 0 \\ \end{array}\right), 
\ \ {\bf e}_3 = \left(\begin{array}{c} 0 \\ 0 \\ 1 \\ \end{array}\right),
\label{vectors3}
\eeq
implying that the principal axes in the $xy$ plane are rotated, relative 
to the $x$ and $y$ axes, through an angle $3/2\gamma t$, which we can 
interpret as the angle between the average orientation of the domain 
structure and the flow direction.

The three length scales are all distinct, though $L_\perp$ and $L_3$ grow 
in the same way, and coarsening continues indefinitely -- the system does not 
approach a stationary state. The matrix elements $M_{ab}$ grow in the way 
expected if a naive form of scaling holds: $M_{11} = L_\parallel^2$, 
$M_{12} \sim L_\parallel L_\perp$, $M_{22} \sim L_\perp^2$, and 
$M_{33} = L_3^2$. This means that dynamical scaling holds, and the 
scaling variables can be taken to be $x/L_\parallel$, $y/L_\perp$, 
and $z/L_3$. The same simple structure does not, however, hold in $d=2$.

{\bf The case} ${\bf d=2}$. For $d=2$ the self-consistency problem is more 
tricky, because the quantity $1-D_{22}(t) = D_{11}(t)$ tends to zero as 
$t \to \infty$. For $d=2$ the integrals (\ref{Dint1}) can be 
evaluated exactly. Making the assumptions, to be verified subsequently, 
that asymptotically $M_{11} \gg M_{12} \gg M_{22}$, that 
${\rm Tr}\,M  \gg \sqrt{\det M} \gg M_{22}$, and that $D_{12} \sim 1/t$, 
one can derive the following self-consistent equation for $D_{11}(t)$:  
\beq
D_{11}(t)= \frac{1}{\gamma t^2}\left(
\frac{\int_0^t dt'\,t'^2 D_{11}(t')}{\int_0^t dt'\,D_{11}(t')}
\right)^{1/2} \ ,
\eeq
with asymptotic solution
\beq 
D_{11}(t) = \frac{1}{2\gamma t\sqrt{\ln \gamma t}}.
\eeq
Using this result, the asymptotic results for the matrix elements, and the 
determinant,  of the correlation matrix $M$ are obtained as
\beqa
M_{11}(t) & =  & 4 \gamma\, t^2\sqrt{\ln \gamma t}  \non \\
M_{12}(t) & =  & 4 t \, \sqrt{\ln \gamma t} \non \\
M_{22}(t) & =  & (4/\gamma)\,\sqrt{\ln \gamma t} \non \\
\det M (t) & = & 4 t^2 \ , 
\label{det}
\eeqa
while $D_{12} = -1/\gamma t$. These results confirm, {\em a posteriori}, 
the assumptions made in their derivation, i.e.\  the solution is 
self-consistent. The asymptotic results for the $M_{ab}$ seem to imply 
$\det M=0$, in contradiction to (\ref{det}). To obtain (\ref{det}) one 
has to keep subdominant contributions to the $M_{ab}$. [Note that 
$\det M = \det R$ from (\ref{M})].

The characteristic length scales are given, as before, by the square roots 
of the eigenvalues of $M$. The eigenvalues of any $2\times2$ matrix can be 
expressed as 
$\lambda_{\pm} = [{\rm Tr}\,M \pm \sqrt{({\rm Tr}\,M)^2 - 4 \det M}]/2$. 
Using ${\rm Tr}\,M \gg \sqrt{\det M}$ we obtain $\lambda_+ = {\rm Tr}\,M$ 
and $\lambda_- = \det M/({\rm Tr}\,M)$, whence
\beq
L_\parallel = 2\sqrt{\gamma}t\,(\ln \gamma t)^{1/4},\ \ \ 
L_\perp = \frac{1}{\sqrt{\gamma}(\ln \gamma t)^{1/4}}\ .
\eeq
The corresponding eigenvectors are 
\beq
{\bf e}_\parallel = 
\left(\begin{array}{c} 1 \\ \frac{1}{\gamma t} \\ \end{array} \right),\ \ \  
{\bf e}_\perp = 
\left(\begin{array}{c} -\frac{1}{\gamma t} \\ 1 \\ \end{array} \right),
\label{vectors2}
\eeq 
giving a tilt angle $1/\gamma t$ between the domain orientation and the flow 
direction.

The scale area in two dimensions is $L_\parallel L_\perp = 2t$, independent 
of $\gamma$. This is the same result as the zero-shear case, where 
$D_{ab}= \delta_{ab}/d$ implies $M_{ab} = 2t\,\delta_{ab}$ for $d=2$, i.e.\ 
$L(t) = \sqrt{2t}$. This result is special to $d=2$ and can be understood 
as follows. For an isolated domain, the rate of change of the area enclosed 
by the domain boundary is  
$dA/dt = \oint dl\,v_n = \oint dl\,({\bf u} - \nabla)\cdot{\bf n}$ from 
(\ref{AC}). The second term is a topological invariant, equal to $-2\pi$ 
from the Gauss-Bonnet Theorem, while the first term is equal to 
$\int_A d^2x\,\nabla \cdot {\bf u}$, which vanishes for any 
divergence-free shear flow. While the $\gamma$-independence of $dA/dt$ has 
been proved only for closed loops of domain wall, we expect a similar result 
to hold for the scale area, i.e. $d(L_\parallel L_\perp)/dt = {\rm const}$. 
It is very encouraging that the OJK approximation captures this 
feature of the $d=2$ coarsening. There is no equivalent result in $d=3$ 
because the surface integral $\int dS\,\nabla \cdot {\bf n}$ is no longer 
a topological invariant.

The scaling is nontrivial in $d=2$ because, although $M_{11} = 
L_\parallel^2$, consistent with naive scaling, the corresponding results 
$M_{12} \sim L_\parallel L_\perp$, $M_{22} \sim L_\perp^2$, found in $d=3$ 
no longer hold in $d=2$. This is associated with the fact that the 
leading-order contribution to $\det M$ vanishes. The consequence is that 
scaling only holds when referred to the unique scaling axes 
(\ref{vectors2}), which are themselves time-dependent.

The most interesting and suggestive feature of the $d=2$ result is that 
$L_\perp$ tends to zero as $t \to \infty$. Since our treatment is based 
on the `thin wall' limit, in which domain walls are treated as having 
zero width, it will break down when $L_\perp$ becomes comparable with 
the width, $\xi$, of the walls, at which point we conjecture 
that domains will break, possibly arresting the coarsening. This  
can be tested by experiments on twisted nematic liquid crystals. 
In $d=3$, the present work provides strong evidence that, at least for 
the nonconserved scalar field considered here, the coarsening state 
proceeds indefinitely. In this respect it is interesting that, in their 
$d=3$ simulations (including both conservation of the order parameter 
and hydrodynamics), Cates {\em et al.} \cite{Cates} found no evidence 
for a steady state structure emerging that is independent of the system 
size, i.e.\ observed steady states could be attributed to finite size 
effects.  

We thank Peter Sollich for a useful discussion. This work was supported 
by EPSRC (UK) under grant GR/L97698.

\end{multicols}


\begin{references}
\bibitem{Review} A. J. Bray, Adv.\ Phys.\ {\bf 43}, 357 (1994), 
and references therein. 
\bibitem{OnukiRev}A. Onuki, J. Phys.: Condens. Matter 
{\bf 9}, 6119 (1997), and references therein. 
\bibitem{Beysens}C. K. Chan, F. Perrot, and D. Beysens, 
Phys.\ Rev.\ A {\bf 43}, 1826 (1991); 
A. H. Krall, J. V. Sengers, and K. Hamano, 
Phys.\ Rev.\ Lett.\ {\bf 69}, 1963 (1992); 
T. Hashimoto, K. Matsuzaka, E. Moses, and A. Onuki, 
Phys.\ Rev.\ Lett.\ {\bf 74}, 126 (1995); 
J. L\"auger, C. Laubner, and W. Gronski, Phys.\ Rev.\ Lett.\ 
{\bf 75}, 3576 (1995).
\bibitem{Rothman}D. H. Rothman, Phys.\ Rev.\ Lett.\ {\bf 65}, 3305 (1990);  
P. Padilla and S. Toxvaerd, J. Chem.\ Phys.\ {\bf 106}, 2342 (1997); 
A. J. Wagner and J. M. Yeomans, Phys.\ Rev.\ E {\bf 59}, 4366 (1999). 
\bibitem{Ohta} T. Ohta, H. Nozaki, and M. Doi, Phys.\ Lett.\ A {\bf 145}, 
304 (1990); J. Chem.\ Phys.\ {\bf 93}, 2664 (1991). 
\bibitem{Cates} M. E. Cates, V. M. Kendon, P. Bladon, and J.-C. Desplat, 
Faraday Discuss.\ {\bf 112}, 1 (1999). 
\bibitem{Rapapa}N. P. Rapapa and A. J. Bray, Phys.\ Rev.\ Lett.\ {\bf 83}, 
3856 (1999); F. Corberi, G. Gonnella, and A. Lamura, Phys.\ Rev.\ Lett.\ 
{\bf 81}, 3852 (1998). 
\bibitem{CZBH} A. Coniglio and M. Zannetti, Europhys.\ Lett.\ 
{\bf 10}, 575 (1989); A. J. Bray and K. Humayun, Phys.\ Rev.\ Lett.\ 
{\bf 68}, 1559 (1992).
\bibitem{OJK} T. Ohta, D. Jasnow, and K. Kawasaki, Phys.\ Rev.\ Lett.\ 
{\bf 49}, 1223 (1982). 
\bibitem{Orihara} H. Orihara and Y. Ishibashi, J. Phys.\ Soc.\ Jpn.\ 
{\bf 55}, 2151 (1986). 
\bibitem{AC} S. M. Allen and J. W. Cahn, Acta Metall.\ {\bf 27}, 1085 (1979). 
\bibitem{CB} A. Cavagna and A. J. Bray, in preparation. 
 
\end{references}
\end{document}